\documentclass[eqsecnum,aip,jmp,reprint]{revtex4-1}
\usepackage[T1]{fontenc}	
\usepackage{newpxtext,newpxmathfixed,mathrsfs}
\usepackage{mathrsfs}
\usepackage{amsmath,amssymb}
\usepackage[sans]{dsfont}
\usepackage{hyperref}
\usepackage{colordvi}
\begin{document}
\begin{titlepage}

\title{Spin in Schr\"odinger-quantized Pseudoclassical Systems}
         
\author{Theodore J. Allen} 
\email{tjallen@hws.edu}
\affiliation{Department of Physics, Hobart and William Smith Colleges \\
Geneva, New York 14456 USA }
\date{\today}
\thispagestyle{empty}
\begin{abstract}
We examine the construction of the spin angular momentum in systems with
pseudoclassical Grassmann variables. In constrained systems there are many
different algebraic forms for the dynamical variables that will all agree
on the constraint surface. For the angular momentum, a particular form of
the generators is preferred, which yields superselection sectors of
irreducible $\mathfrak{spin}^c(n)$ representations rather than reducible
$\mathfrak{so}(n)$ representations when quantized in the Schr\"odinger
realization.
\end{abstract}
\maketitle

\end{titlepage}

\section{Introduction}\label{sec:introduction}
Quantization of the pseudoclassical
actions\cite{Berezin:1975md,Berezin:1976eg,Casalbuoni:1976tz,Brink:1976uf,Barducci:1976qu,Brink:1976sz,Balachandran:1976ya},
introduced more than forty years ago to describe spinning particles, has
been done by the path integral method and by canonical quantization
directly on the reduced phase space.  Quantization in the Schr\"odinger
picture was partially worked out by Barducci, Bordi, and
Casalbuoni\cite{Bordi:1980aj,Barducci:1980rx} but, as far as we are aware,
the full details have been worked out only recently\cite{Allen:2015lma}.

In the standard reduced phase space approach, the phase space has
rotational covariance, which must be broken by choosing a particular
splitting into coordinates and momenta---a ``polarization''---in order to
use the Schr\"odinger realization.  Because the actions are first-order in
velocities, in order to quantize without reducing the phase space first we
must use Dirac's constrained Hamiltonian quantization, but in exchange we
gain a coordinate space that is rotationally covariant. In the reduced
phase space approach, the Noether angular momentum directly gives the
correct spin, while in the Dirac quantization, the Noether angular momentum
is ambiguous and one particular form must be chosen to obtain the correct
spin.

Construction of angular momentum as differential operators on functions of
Grassmann variables is not new.  Manko\v{c}-Bor\v{s}tnik gave such a
construction,\cite{MankocBorstnik:1993ia,MankocBorstnik:1995zu} though not
in the context of the Schr\"odinger quantization of a constrained
pseudoclassical mechanical system. What is new here is to examine angular
momentum in the context of the constrained Schr\"odinger quantization of
the simplest pseudoclassical systems.

In the Schr\"odinger realization, the full state space splits into
orthogonal physical and ghost sectors that have positive-definite and
negative-definite norms, respectively. Within those sectors, for dimensions
greater than two, there are isomorphic superselection sectors, each of
which corresponds to quantization directly on the reduced phase space and
forms an irreducible representation of $\mathfrak{spin}^c(n)$.

The wave functions in the Schr\"odinger realization thus correspond to
spinorial states, and can be directly
mapped\cite{MankocBorstnik:1999th,Allen:2015lma} to K\"ahler
fermion\cite{Ivanenko:1928zf,Kahler:1960zz,Kahler:1961as,Kahler:1962id}
differential form-valued wave functions.

In the following, we examine how the $\mathfrak{so}(n)$ algebra that acts
on the physical state space becomes $\mathfrak{spin}^c(n)$. We use the
Einstein summation convention that repeated indices are summed. Because our
metric is the unit matrix, there is no distinction between upper and lower
indices but we raise or lower them for notational convenience and to make
the generalization to indefinite metrics nearly immediate.

\section{Pseudoclassical action}\label{sec:pseudoclassicalaction}
We consider systems of $n$ anticommuting variables $\xi^i$ described by the
rotationally invariant action
\begin{align}\label{eq:theaction}
S & =  \int d\tau\, \Big[ \frac{i}{2}\xi_i\dot\xi^i - H(\xi) \Big] \, ,
\end{align}
as a Hamiltonian system with constraints, in the sense of Dirac. The
canonical Poisson brackets of two phase space functions, $A(\pi,\xi)$ and
$B(\pi,\xi)$ is given by
\begin{equation}
\big\{ A, B \big\} = {\partial^R A\over \partial\pi_i} {\partial^L B\over
  \partial\xi^i} + {\partial^R A\over \partial\xi^i} {\partial^L B\over
  \partial\pi_i}\, ,
\end{equation}
where the $R$ and $L$ superscripts denote whether the derivative is to be
taken from the right or the left.

\section{Dirac Constraint analysis}\label{eq:constraints}
Because the action (\ref{eq:theaction}) is first-order in velocities, there
are Dirac constraints\cite{Dirac:1964:LQM,Hanson:1976cn},
\begin{align}
\varphi^i & =  \pi^i - \frac{i}{2} \xi^i  \approx 0 \, . \label{eq:seccls} 
\end{align} 
These constraints have constant, non-vanishing Poisson brackets with each
other
\begin{align}\label{eq:secclsPBs}
\Delta^{ij} & = \big\{\varphi^i,\varphi^j\big\} = -i\delta^{ij} \, , 
\end{align}
and so are second-class. The matrix (\ref{eq:secclsPBs}) has inverse
\begin{align}\label{eq:secclsPBinverse}
\Delta^{-1}_{ij} & =  i\delta_{ij} \, .
\end{align}
The dynamical system can be reduced to the phase space defined by the
constraints (\ref{eq:seccls}) with the (ortho-)sym\-plectic form given by
the Dirac bracket,
\begin{equation}
\big\{ A , B \big\}_D = \big\{ A , B \big\} - \big\{ A , \varphi^i \big\}
\Delta^{-1}_{ij} \big\{ \varphi^j , B \big\}\, ,
\end{equation} 
which allows second-class constraints to be taken to zero strongly because
they will have vanishing Dirac brackets with any dynamical quantity,
\begin{equation}
\big\{ A , \varphi^i \big\}_D \equiv 0 \, .
\end{equation}
The Dirac brackets satisfy the same (graded) Jacobi identity as the Poisson
brackets.

Instead of using Dirac brackets, dynamical variables can be replaced by
their ``primed'' versions\cite{Hanson:1976cn}
\begin{equation}
A^\prime = A - \big\{ A , \varphi^i \big\} \Delta^{-1}_{ij} \varphi^j , 
\end{equation}
and the original Poisson bracket can be kept, at the possible cost of
having certain relations holding only on the constraint surface, rather
than the full phase space.  The Poisson brackets of primed variables
satisfy
\begin{equation}
\big\{ A^\prime , B^\prime \big\} \approx \big\{ A^\prime , B \big\} \approx \big\{ B , A^\prime \big\} \approx \big\{ A , B \big\}_D\, , 
\end{equation}
where we have used Dirac's ``weak equality,'' $\approx$, to denote
quantities whose difference vanishes on the constraint surface defined by
the constraints, in this case Eqs.~(\ref{eq:seccls}).

We have previously analyzed the system described by the action
(\ref{eq:theaction}) in detail\cite{Allen:2015lma} using the primed
variables
\begin{equation}\label{eq:primed_variables}
\begin{split}
  \xi^{\prime\, i} & \equiv \xi^i - \big\{ \xi^i, \varphi^a \big\}
  \Delta^{-1}_{ab} \varphi^b = \xi^i - i \varphi^i \\
  & = \frac{1}{2}\xi^i - i\pi^i\, , \\
  \pi^{\prime\, i} & \equiv \pi^i - \big\{ \pi^i, \varphi^a \big\}
  \Delta^{-1}_{ab} \varphi^b = \pi^i -
  \frac{1}{2}\varphi^i \\
  & = \frac{1}{2}\pi^i + \frac{i}{4}\xi^i = \frac{i}{2}\xi^{\prime\, i} .
\end{split}
\end{equation}
These variables have strongly vanishing Poisson brackets with the
second-class constraints,
\begin{equation}\label{eq:HRTxichi}
\begin{split}
\big\{ \xi^{\prime\, i} , \varphi^j \big\} & = 0 \, , \\
\big\{ \pi^\prime_i , \varphi^j \big\} & = 0 \, .
\end{split}
\end{equation}
After quantization, the Poisson brackets become anticommutators and the set
of $\hat\xi^{\prime\, i}$ then generate a Clifford
algebra.\cite{Allen:2015lma}

\section{Angular momentum}\label{sec:constants}
The conserved angular momentum of the system follows from Noether's theorem
applied to infinitesimal rotations of the variables
\begin{align}\label{eq:infrot}
\delta_\omega \xi^i  & =  \omega^i{}_{j} \, \xi^j \, , \quad \omega_{ji} = -\omega_{ij} \, .
\end{align}
When the action (\ref{eq:theaction}) is invariant under infinitesimal
rotations, the definition and conservation of the spin angular momentum
${S}\mathstrut^{ij}$ follow from the equations of motion and the invariance
of the action under (\ref{eq:infrot}),
\begin{align}\label{eq:noether1}
\delta_\omega S & = \Delta\left({\partial^R L \over \partial \dot
\xi^i} \delta_\omega\xi^i \right) = \frac12 \omega_{ij} \Delta\, {S}\mathstrut^{ji} = 0\, ,
\end{align}
where $\partial^R/\partial \dot\xi^i$ denotes the derivative acting from
the right, and $\Delta$ denotes the difference in values between final and
initial times.

The angular momentum from (\ref{eq:noether1}) is
\begin{align}\label{eq:NoetherSpin}
{S}^{ij} & = - \xi^{[i}\pi^{j]} \equiv - \xi^i\pi^j + \xi^j \pi^i \, .
\end{align}
Even for non-invariant actions, the angular momentum still generates
rotations of the canonical variables,
\begin{equation}
\delta_\omega z = \frac{1}{2}\omega^{ji} \big\{ z, S_{ij} \big\} \, .
\end{equation}

\section{Weakly equivalent angular momenta}
First, we note the constraints (\ref{eq:seccls}) are vectors, and we have
the Poisson brackets
\begin{align}\label{eq:vectornatureconstraints}
\big\{S^{ij}, \varphi^k \big\} = \delta^{ik}\varphi^{j} - \delta^{jk}\varphi^i = \delta^{k[i}\varphi^{j]} \approx 0 \, .
\end{align}
It is therefore consistent on the constraint surface to use the Poisson
bracket rather than the Dirac bracket with the angular momentum.

{}From (\ref{eq:vectornatureconstraints}) we can calculate the
Hanson-Regge-Teitelboim primed version of the angular momentum,
\begin{equation}\label{eq:sprime}
\begin{split}
S^{\prime\, i j} & = S^{ij} - \big\{S^{ij},\varphi^a \big\}\Delta^{-1}_{ab}\varphi^b \\
& = S^{ij} + i \varphi^{[i}\varphi^{j]} \, . \\
\end{split}
\end{equation}
We find the Poisson brackets 
\begin{align}\label{eq:primedLbrackets}
\big\{ S^{\prime\, ij},\varphi^k \big\} & = \big\{S^{ij},\varphi^k \big\} + i \big\{\varphi^{[i} \varphi^{j]}, \varphi^k \big\} \nonumber \\
&= \delta^{k[i} \varphi^{j]} + 2i \varphi^{[i}(-i \delta^{j]\,k}) \\
&= - \delta^{k[i} \varphi^{j]} \approx 0\, , \nonumber 
\end{align}
which show that the primed version of the angular momentum
(\ref{eq:sprime}) is no better than the original canonical version
(\ref{eq:NoetherSpin}) in allowing the use of Poisson rather than Dirac
brackets.

\section{Weak and Strong Lie Algebras}
Noting that under the primed generators ${S}^{\prime\,ij}$, the constraints
transform (\ref{eq:primedLbrackets}) contravariantly rather than
covariantly, as they do under the $S^{ij}$, we now consider the
one-parameter family of weakly equal forms of the angular momentum
\begin{equation}\label{eq:sigmagenerators}
\Sigma^{ij}_\epsilon = S^{ij} + i\epsilon \varphi^{[i}\varphi^{j]}\, ,
\end{equation}
and examine whether their Lie algebras close strongly or only weakly.

The generators $J^{ij}$ of $\mathfrak{so}(n)$ (and also
$\mathfrak{spin}^c(n)$), satisfy the Poisson bracket algebra
\begin{equation}\label{eq:son}
\big\{ J^{ij}, J^{mn} \big\} = -\, \delta^{jm}J^{in} + \delta^{jn} J^{im} -
\delta^{in} J^{jm} + \delta^{im} J^{jn} \, .
\end{equation}
It is straightforward to check that the canonical Noether generators
(\ref{eq:NoetherSpin}) satisfy the relations (\ref{eq:son}) strongly.

{}From the Poisson brackets
\begin{align}
\big\{ S^{ij}, S^{mn} \big\} & = - \delta^{jm}S^{in} + \delta^{jn} S^{im} - \delta^{in} S^{jm} + \delta^{im} S^{jn}  \, , \nonumber
\end{align}
\begin{equation}
\begin{split}
\big\{S^{ij}, \varphi^{[m}\varphi^{n]} \big\} = & - \delta^{jm}\,
\varphi^{[i} \varphi^{n]} + \delta^{jn}\, \varphi^{[i}\varphi^{m]} \\ & -
\delta^{in} \varphi^{[j}\varphi^{m]} + \delta^{im} \varphi^{[j}\varphi^{n]}
\, ,
\end{split}
\end{equation}
\begin{equation}
\begin{split}
\big\{ \varphi^{[i} \varphi^{j]}, S^{mn} \big\} & = - \big\{S^{mn},
\varphi^{[i} \varphi^{j]} \big\} \nonumber \\ & = \big\{S^{ij},
\varphi^{[m}\varphi^{n]} \big\} \, ,
\end{split}
\end{equation}
and
\begin{equation}\label{eq:chichi}
\begin{split}
\big\{i\varphi^{[i}\varphi^{j]}, i\varphi^{[m}\varphi^{n]} \big\} & =  2 \delta^{jm}\, i \varphi^{[i} \varphi^{n]} - 2\delta^{jn}\, i\varphi^{[i}\varphi^{m]}  \\
             & + 2\delta^{in} i \varphi^{[j}\varphi^{m]} -2 \delta^{im} i\varphi^{[j}\varphi^{n]} \, , 
\end{split}
\end{equation}
we can determine the Poisson brackets of the quantities (\ref{eq:sigmagenerators}) as
\begin{equation}\label{eq:sigmason}
\begin{split}
\big\{ \Sigma^{ij}, \Sigma^{mn} \big\} = 
             & -\delta^{jm}\,\left(S^{in} + 2(1-\epsilon) i\epsilon \varphi^{[i} \varphi^{n]} \right) \\ 
             & + \delta^{jn}\,\left(S^{im} + 2(1-\epsilon) i\epsilon\varphi^{[i} \varphi^{m]} \right) \\
             & - \delta^{in}\,\left(S^{jm} + 2(1-\epsilon) i\epsilon\varphi^{[j} \varphi^{m]} \right) \\
             & + \delta^{im}\,\left(S^{jn} +  2(1-\epsilon) i\epsilon\varphi^{[j} \varphi^{n]} \right)\, ,  
\end{split}
\end{equation}
and see that the generators (\ref{eq:sigmagenerators}) will satisfy the Lie
algebra relations (\ref{eq:son}) strongly only for
\begin{equation}
\epsilon = 0\quad {\rm or}\quad \epsilon = \frac12 \, . 
\end{equation}
For other values of $\epsilon$ the algebra (\ref{eq:son}) is only satisfied
weakly.
We observe that the Poisson bracket of the generator $\Sigma^{ij}$ and the
constraint $\varphi^k$,
\begin{equation}
\begin{split}
\big\{ \Sigma^{ij}, \varphi^k \big\}  & =  \big\{S^{ij},\varphi^k \big\} + i\epsilon \big\{\varphi^{[i} \varphi^{j]}, \varphi^k \big\} \\
&= \delta^{k[i} \varphi^{j]} + 2i\epsilon \varphi^{[i}(-i \delta^{j]\,k}) \\
&= (1 - 2\epsilon) \delta^{k[i} \varphi^{j]} \approx 0\, ,
\end{split}
\end{equation}
also vanishes strongly if and only if $\epsilon = 1/2$. It's also
instructive to note that when $\epsilon = 1/2$,
\begin{equation}\label{eq:sigmainprimes}
\Sigma^{ij} = - \xi^{\prime\, [i} \pi^{\prime\, j]} = - \frac{i}{2} \xi^{\prime\, [i} \xi^{\prime\, j]} \, ,
\end{equation}
which immediately implies $\big\{ \Sigma^{ij} , \varphi^k \big\}
\equiv 0$.

Eq.~(\ref{eq:chichi}) shows that the quantities
$-\frac{i}{2}\varphi^{[i}\varphi^{j]}$ also satisfy the Poisson bracket
relations (\ref{eq:son}), and with Eq.~(\ref{eq:sigmainprimes}), implies
that the canonical $\mathfrak{so}(n)$ generators (\ref{eq:NoetherSpin}) are
a sum of two independent, commuting $\mathfrak{spin}^c(n)$ generators,
\begin{equation}\label{eq:halfnhalf}
S^{ij} = - \frac{i}{2} \xi^{\prime\, [i} \xi^{\prime\, j]} - \frac{i}{2} \varphi^{[i} \varphi^{j]} \, .
\end{equation}

\section{Quantum Mechanics}
The Schr\"odinger realization\cite{Allen:2015lma} of quantum mechanics for
theories described by the action (\ref{eq:theaction}) with $n$ Grassmann
variables consists of replacement of the canonical variables by the
operators
\begin{equation}
\begin{split}
\hat{\xi}_i \psi(\xi) & = \xi_i\, \psi(\xi)\, , \\
\hat{\pi}_i \psi(\xi) & = i{\partial^L \over \partial \xi_i}\, \psi(\xi)\, , 
\end{split}
\end{equation}
that act on wave functions $\psi(\xi)$ of all $n$ Grassmann variables. The
Schr\"odinger inner product is given by the integral over the full
configuration space,
\begin{align}\label{eq:SchrodNorm}
  \langle \phi|\psi \rangle & = i^{\lfloor \frac{n}{2}\rfloor} \int \phi^* \psi\, d\xi_1 d\xi_2 \ldots d\xi_n = \langle \psi | \phi \rangle^* \> , \phantom{\Bigg|}
\end{align}
under which the coordinates variables $\hat\xi_i$ are self-adjoint, and the
momentum variables $\hat\pi_i$ are anti-self-adjoint.

The second-class constraints $\hat\varphi_i \approx 0$ are imposed by the
generalized Gupta-Bleuler condition that all physical matrix elements of
the constraints vanish,
\begin{align}
\langle \phi_{\rm phys} | \hat\varphi_i | \psi_{\rm phys} \rangle = 0 \, .
\end{align}
This condition splits the full state space into two orthogonal subspaces:
positive norm physical states and negative norm unphysical (ghost)
states. Note that if the opposite sign is taken in (\ref{eq:SchrodNorm}),
the physical and ghost states are interchanged. The constraints map
physical states into ghost states, and vice-versa.\cite{Allen:2015lma}

If all dynamical variables have weakly vanishing commutators or
anti-commutators with the constraints, then the constraints can be taken to
be strongly zero and the Hilbert space restricted to physical
states. Replacing the Grassmann operators $\hat\xi_i$ by $\hat\xi^\prime_i$
and $\hat\pi_i$ by $\hat\pi^\prime_i$ in all dynamical variables built from
them will accomplish this.

\subsection{Two dimensions}
In a system with just two anticommuting coordinates, the physical (positive
norm) states are spanned by the orthonormal states\cite{Allen:2015lma}
\begin{equation}\label{eq:fullphysbasis}
\begin{split}
|{0}\rangle     &= e^{\frac{i}{2}\xi_1\xi_2} = 1 + \frac{i}{2}\xi_1\xi_2 \>  , \\
|{1}\rangle     &= \frac{1}{\sqrt2}\left(\xi_1 + i\xi_2\right)\, e^{\frac{i}{2}\xi_1\xi_2} = \frac{1}{\sqrt2}\left(\xi_1 + i\xi_2\right) \>  ,
\end{split}
\end{equation}
while the ghost (negative norm) states are spanned by the orthonormal states
\begin{equation}\label{eq:fullghostbasis}
\begin{split}
|\bar{0}\rangle &= e^{-\frac{i}{2}\xi_1\xi_2} = 1 - \frac{i}{2}\xi_1\xi_2  \> , \\
|\bar{1}\rangle &= \frac{1}{\sqrt2}\left(\xi_1 - i\xi_2\right)\,e^{-\frac{i}{2}\xi_1\xi_2} = \frac{1}{\sqrt2}\left(\xi_1 - i\xi_2\right)  \>  .
\end{split}
\end{equation}

Denoting the general state of the system by
\begin{equation}
\begin{pmatrix} a \\ b \\ c \\ d \end{pmatrix} = a |0\rangle + b | 1 \rangle + c | \bar 0 \rangle + d |\bar 1 \rangle\, , 
\end{equation}
we can write the matrix representation of the canonical position and
momentum operators on the general state as
\begin{equation}
\begin{split}
\hat\xi_1 &= \frac{1}{\sqrt2} \begin{pmatrix} \sigma_1 & -i \sigma_2 \\ -i\sigma_2 & \sigma_1 \end{pmatrix} \, ,\quad 
 \hat\xi_2  = \frac{1}{\sqrt2} \begin{pmatrix} - \sigma_2 & - \sigma_2 \\ \sigma_2 & \sigma_2 \end{pmatrix} \, , \\ 
%
\hat\pi_1 & = \frac{i}{2\sqrt2}\begin{pmatrix} \sigma_1 & i \sigma_2 \\ i\sigma_2 & \sigma_1 \end{pmatrix} \, , \quad
\hat\pi_2  = \frac{1}{2\sqrt2}\begin{pmatrix} -i \sigma_2 & i \sigma_2 \\ -i\sigma_2 & i \sigma_2 \end{pmatrix}\, , 
\end{split}
\end{equation}
where $\sigma_1$ and $\sigma_2$ are standard Pauli matrices.  The diagonal
pieces map physical states to physical states and unphysical to unphysical,
while the off-diagonal pieces map physical to unphysical states or
vice-versa. In this matrix representation, the canonical Noether angular
momentum is
\begin{equation}
\hat S_{12} = \hat\xi_2 \hat\pi_1 - \hat\xi_1\hat\pi_2 = \frac{1}{2} \begin{pmatrix} \mathds{1} - \sigma_3 & 0 \\ 0  & -\mathds{1} + \sigma_3 \end{pmatrix} \, .
\end{equation}
Including the term containing the second-class constraints and with
$\epsilon={1}/{2}$, we find
\begin{equation}
\hat \Sigma_{12} = \hat S_{12} + i\hat\varphi_1 \hat\varphi_2 = \frac{1}{2} \begin{pmatrix} - \sigma_3 & 0 \\ 0  & + \sigma_3 \end{pmatrix} \, .
\end{equation}

\subsection{Three dimensions}
Adding one more anticommuting coordinate to the two-dimensional system
creates two superselection sectors of physical states and two of ghost
states. These are superselection sectors for the Hanson-Regge-Teitelboim
$\hat\xi^\prime_i$ and $\hat\pi^\prime_i$ operators, and are denoted by
unprimed and primed states.\cite{Allen:2015lma}

The positive norm physical states are spanned by the orthonormal basis
\begin{equation}\label{eq:threexiphys}
\begin{split}
 |0\rangle & = \frac{1}{\sqrt[4]2}\Big(1 + \frac{i}{2}\xi_1\xi_2\Big)\left (1 +
 \frac{\xi_3}{\sqrt2}\right) \> ,\\
 |1\rangle & =  \frac{1}{\sqrt[4]8} \Big(\xi_1 + i\xi_2\Big)\left(1 + \frac{\xi_3}{\sqrt2}\right) \> ,\\
 |0^\prime\rangle & =  \frac{1}{\sqrt[4]2}\Big(1 - \frac{i}{2}\xi_1\xi_2\Big) \left(1 -
\frac{\xi_3}{\sqrt2}\right) \> , \\
|1^\prime\rangle & =  \frac1{\sqrt[4]8} \Big(\xi_1 - i\xi_2\Big)\left(1 - \frac{\xi_3}{\sqrt2}\right) \> ,
\end{split}
\end{equation}
while the negative norm ghost states are denoted by barred states and
spanned by the orthogonal anti-normal basis
\begin{equation}
\begin{split}
 |\bar{0}\rangle  & =  \frac{1}{\sqrt[4]2}\Big(1 - \frac{i}{2}\xi_1\xi_2\Big) \left(1 +
 \frac{\xi_3}{\sqrt2}\right) \> , \\  
 |\bar{1}\rangle  & =  \frac1{\sqrt[4]8} \Big(\xi_1 - i\xi_2\Big)\left(1 + \frac{\xi_3}{\sqrt2}\right) \> , \\
 |\bar{0}^\prime\rangle & =  \frac{1}{\sqrt[4]2}\Big(1 + \frac{i}{2}\xi_1\xi_2\Big)
 \left(1 - \frac{\xi_3}{\sqrt2}\right) \> , \\ 
 |\bar{1}^\prime\rangle & =  \frac1{\sqrt[4]8} \Big(\xi_1 + i\xi_2\Big)\left(1 - \frac{\xi_3}{\sqrt2}\right) \>  . 
\end{split}
\end{equation}
In this full basis, where the general state of the system is denoted by
\begin{equation}\label{eq:threebasis}
\begin{pmatrix} a \\ b \\ c \\ d \\ e \\ f \\ g \\ h\end{pmatrix} = a |0\rangle + b | 1 \rangle + c | \bar 0 \rangle + d |\bar 1 \rangle + e | 0^\prime \rangle + f | 1^\prime \rangle + g | \bar{0}^\prime\rangle + h | \bar{1}^\prime \rangle , 
\end{equation}
the coordinate and momentum operators are represented, using the standard
Pauli matrices, by the $8\times8$ matrices

\begin{align} 
\hat\xi_1 & = \frac{1}{\sqrt2} \left(\begin{array}{c c} 
         \begin{array}{c c} 
          \sigma_1 & -i\sigma_2 \\ 
          -i\sigma_2 & \sigma_1 
       \end{array} & \text{\Large 0} \\
      \text{\Large 0} & \begin{array}{c c}
           \sigma_1 & -i \sigma_2 \\
           -i\sigma_2 & \sigma_1
          \end{array} 
     \end{array} 
    \right) \ , 
\end{align}
\begin{align}
\hat\xi_2 & = \frac{1}{\sqrt2} \left(\begin{array}{c c} 
         \begin{array}{c c} - \sigma_2 & - \sigma_2 \\ \sigma_2 & \sigma_2 \end{array} & \text{\Large 0} \\ 
         \text{\Large 0} & \begin{array}{c c} \sigma_2 & \sigma_2 \\ -\sigma_2 & - \sigma_2 \end{array} 
                                     \end{array} \right) \ , 
\end{align}
\begin{align}
\hat\xi_3 & = \frac{1}{\sqrt2} \begin{pmatrix} \sigma_3 & 0 & 0 & \sigma_3 \\ 0 & \sigma_3 & \sigma_3 & 0 \\
                                               0 & -\sigma_3 & -\sigma_3 & 0 \\ -\sigma_3 & 0 & 0 & -\sigma_3 \end{pmatrix} \, , 
\end{align}
\begin{align} 
\hat\pi_1 & = \frac{1}{2\sqrt2} \left(\begin{array}{c c} 
         \begin{array}{c c} 
          i\sigma_1 & - \sigma_2 \\ 
          -\sigma_2 & i \sigma_1 
       \end{array} & \text{\Large 0} \\
      \text{\Large 0} & \begin{array}{c c}
           i\sigma_1 & - \sigma_2 \\
           -\sigma_2 & i\sigma_1
          \end{array} 
     \end{array} 
    \right) \ , 
\end{align}
\begin{align}
\hat\pi_2 & = \frac{1}{2\sqrt2} \left(\begin{array}{c c} 
             \begin{array}{c c} - i\sigma_2 & i \sigma_2 \\ -i\sigma_2 & i\sigma_2 \end{array} & \text{\Large 0} \\ 
             \text{\Large 0} & \begin{array}{c c}  i\sigma_2 & -i\sigma_2 \\ i\sigma_2 & - i\sigma_2 \end{array} 
                                     \end{array} \right) \ , 
\end{align}
\begin{align}   
\hat\pi_3 & = \frac{1}{2\sqrt2} \begin{pmatrix} i \sigma_3 & 0 & 0 & -i\sigma_3  \\ 0 & i\sigma_3 & -i\sigma_3 & 0 \\
                                                0 & i\sigma_3 & -i\sigma_3 & 0 \\  i \sigma_3 & 0 & 0 & -i\sigma_3\end{pmatrix} \, , 
\end{align}
which allow for a more standard and reliable computation tool than Grassmann operators.
                                                
It is straightforward to check that the representation of the canonical
angular momentum algebra of the $\hat{S}_{ij} = -\hat\xi_{[i}\hat\pi_{j]}$,
splits the four physical states into the reducible $\mathfrak{so}(3)$
representation ${\bf 1}\oplus {\bf 3}$, and similarly splits the four ghost
states. The splitting mixes the superselection sectors; the $\hat{\bf S}^2
= 2$, $\hat{S}_{12}=0$ state is
\begin{equation}
| S=1, S_z=0\rangle = \frac{1}{\sqrt2}\left(\mathstrut |0\rangle - |0^\prime\rangle\mathstrut\right)\, , 
\end{equation}
while the singlet state is
\begin{equation}
| S=0, S_z=0\rangle = \frac{1}{\sqrt2}\left(\mathstrut|0\rangle + |0^\prime\rangle\mathstrut\right) \, , 
\end{equation}
and similarly for the ghost sectors. 

Once the constraint modification is made in Eq.~(\ref{eq:sigmagenerators}) with
$\epsilon={1}/{2}$, the $\hat{\Sigma}\mathstrut_{ij}$ exactly
commute with all second-class constraints, making the constraints
scalars under rotation, and the physical and ghost spaces each split up
into superselection sectors as two irreducible $\mathfrak{spin}^c(3)$
representations, ${\bf 2} \oplus {\bf 2}$. This is implied by the relation
Eq.~(\ref{eq:sigmainprimes}), and the fact that the $\hat{\xi}^\prime_i$ do
not mix superselection sectors\cite{Allen:2015lma}.

\subsection{General case}
With $n$ anticommuting coordinates, there are $2^n$ total states, half of
which are physical and half of which are ghost. A ghost state differs from
a physical state by a reflection in one of the $\xi_i$ in its wave
function, which changes the sign of its norm. A rotation cannot change just
one sign, so $SO(n)$ rotations will preserve the sign of the norm.

The $2^{n-1}$ physical states fall into $N_s =
2^{{\lfloor\frac{n-1}{2}\rfloor}}$ superselection sectors of the
$\hat\xi^\prime_i$ operators, and the same is true of the ghost
states. Each superselection sector, physical or ghost, has dimension
$2^{\lfloor \frac{n}{2}\rfloor}$. For odd $n$, these superselection sectors
are irreducible representations of $\mathfrak{spin}^c(n)$. For even $n$,
because the generators (\ref{eq:sigmainprimes}) have even Grassmann parity,
these superselection sectors are a sum of two irreducible spinor
representations of $\mathfrak{spin}^c(n)$ having opposite Grassmann
parity. However, products of odd numbers of the $\hat\xi^\prime_i$
operators will mix them. Additional Grassmann degrees of freedom added to
the theory may prevent these irreducible spinor representations from being
quantum mechanical superselection sectors. From the construction of states
as functions of the Grassmann variables $\xi_i$ given in
Ref.~\onlinecite{Allen:2015lma}, we can see that under the rotations
generated by the $\hat{S}_{ij}$, the superselection sectors will mix,
giving in general a sum of irreducible scalar, vector, and antisymmetric
tensor representations of $SO(n)$ with at most $\lfloor \frac{n}{2}\rfloor$
indices. Explicit constructions of these states is given in Appendix B.

For dimensions smaller than three there is just one superselection
sector. For $n=1$ there is a single scalar state and for $n=2$ there is one
superselection sector consisting of a scalar and a half-vector, while for
$n=5$, for example, the four physical $\mathfrak{spin}^c(5)$ spinor ${\bf
  4}$ superselection sectors will mix under $\mathfrak{so}(5)$ into the
reducible representation ${\bf 1} \oplus {\bf 5 } \oplus {\bf 10 }$, and
for $n=6$, the four physical $\mathfrak{spin}^c(6)$ reducible spinor ${\bf
  8}$ superselection sectors will mix under $\mathfrak{so}(6)$ into the
reducible representation ${\bf 1} \oplus {\bf 6 } \oplus {\bf 15 } \oplus
{\bf 10}$.  The same is true for the ghost sectors.

In constructing isospin operators to describe relativistic particles
interacting with a non-Abelian gauge field, Balachandran \textit{et al.\/}\
examined constructions of isospin for any representation of an arbitrary
gauge group in terms of both commuting and anticommuting
variables.\cite{Balachandran:1976ya} Two of these constructions are
relevant to our construction of spin from the fundamental representation of
$\mathfrak{so}(n)$.  When the spin is constructed---\textit{e.g.\/}, in a
reduced phase space quantization---from an abstract \textit{n}-dimensional
Clifford algebra as in Eq.~(\ref{eq:sigmainprimes}), a single
$\mathfrak{spin}^c(n)$ representation results.\cite{Allen:2015lma} Because
the Schr\"odinger realization constructs the Clifford algebra generators
$\hat\xi^{\prime\,i}$ in Eq.~(\ref{eq:primed_variables}) as operators on
wave functions of all $n$ of the Grassmann coordinates, repeated
$\mathfrak{spin}^c(n)$ representations occur as identical superselection
sectors.\cite{Allen:2015lma} A consistent quantization can be obtained by
restricting to just one of these superselection sectors, although
restricting to more than one superselection sector is also consistent, and
in the language of K\"ahler fermions has been proposed as the origin of
fermion families.\cite{Banks:1982iq} When the angular momentum is
constructed from two \textit{n}-dimensional Clifford algebras, as in the
decomposition of the Noether angular momentum in Eq.~(\ref{eq:halfnhalf}),
one obtains a $\mathfrak{spin}^c(2n)$
representation\cite{Balachandran:1976ya} that reduces under
$\mathfrak{so}(n)$. In Appendix B we give a tensor construction for the
reduction under $\mathfrak{so}(n)$, different from the method given in
Ref.~\onlinecite{Balachandran:1976ya}.

\section{Conclusions}
The unique generators of the family (\ref{eq:sigmagenerators}) that have
vanishing Poisson brackets with the constraints become the generators of
$\mathfrak{spin}^c(n)$ upon quantization and the physical state space
splits up into superselection sectors that are identical
$\mathfrak{spin}^c(n)$ representations. A consistent quantization can be
obtained by restricting physical states to one or more of these
superselection sectors.

The unique generators of that same family (\ref{eq:sigmagenerators}) under
which the constraints transform as vectors become the generators of
$\mathfrak{so}(n)$ upon quantization and the physical
$\mathfrak{spin}^c(n)$ superselection sectors mix to become a sum of
antisymmetric tensor $\mathfrak{so}(n)$ irreducible representations. One
cannot generally make a useful quantization using the Noether form of the
angular momentum because the Grassmann position and momentum operators must
be modified by the Hanson-Regge-Teitelboim procedure so that they do not
mix physical and ghost states; mixing of ghost and physical states would
ruin the consistent removal of the ghost states from the theory. With the
modified $\xi^\prime$ and $\pi^\prime$ operators, the natural states are in
$\mathfrak{spin}^c(n)$ superselection sectors, which do not mix under the
$\xi^\prime$, unlike the $\mathfrak{so}(n)$ irreps. To use the Noether form
of the angular momentum, the full $2^{n-1}$-dimensional
$\mathfrak{spin}^c(2n)$ physical state space, meaning all the
$\mathfrak{so}(n)$ scalar, vector, and antisymmetric tensor states, would
have to be kept. Such a quantization does not seem to have relevance to
physics.

We also found, in the quantum version of Eq.~(\ref{eq:halfnhalf}), that the
canonical $\mathfrak{so}(n)$ generators (\ref{eq:NoetherSpin}) are the sum
of two independent commuting $\mathfrak{spin}^c(n)$ generators:
\begin{equation}\label{eq:angmom}
\hat{S}^{ij} = -\hat\xi^{[i} \hat\pi^{j]} = - \frac{i}{2} \hat\xi^{\prime\,
  [i} \hat\xi^{\prime\, j]} - \frac{i}{2} \hat\varphi^{[i} \hat\varphi^{j]}
\, .
\end{equation}
After setting the constraints $\hat\varphi_i$ strongly to zero and
restricting the Hilbert space to the physical Hilbert space, the Noether
$\mathfrak{so}(n)$ angular momentum generators (\ref{eq:angmom}) become the
$\mathfrak{spin}^c(n)$ generators, Eq.~(\ref{eq:spincngens}).

The selection of (\ref{eq:sigmagenerators}) with $\epsilon={1}/{2}$,
\begin{equation}\label{eq:spincngens}
\hat\Sigma^{ij} = - \hat\xi^{[i}\hat\pi^{j]} + \frac{i}{2} \hat\varphi^{[i}\hat\varphi^{j]}  = - \frac{i}{2} \hat\xi^{\prime\, [i} \hat\xi^{\prime\, j]} \, , 
\end{equation}
as the correct form of the spin angular momentum depends on only two
things.  One is that the second-class constraints $\varphi^i \approx 0$
transform as vectors under the canonical Noether angular momentum
(\ref{eq:NoetherSpin}) and the other is the specific constant value of the
rotationally invariant Poisson brackets of the constraints given in
Eq.~(\ref{eq:secclsPBs}). Our results thus apply to somewhat more general
actions than the simple action (\ref{eq:theaction}) that we have considered
here.

\section*{Acknowledgments}
We thank an anonymous referee for pointing out Ref.~\onlinecite{Balachandran:1976ya}.

\section*{Data availability statement}
Data sharing not applicable – no new data generated.

\appendix
\section{Operator representations in three dimensions}
We gather here for reference the matrix representations for various
quantities in the basis (\ref{eq:threebasis}), expressed as tensor products
of $2\times 2$ matrices.  The canonical coordinates are given by
\begin{align}
\sqrt2\, \hat\xi_1 & = \mathds{1} \otimes \mathds{1} \otimes \sigma_1 - i \mathds{1} \otimes \sigma_1 \otimes \sigma_2 \, ,\nonumber \\
\sqrt2\, \hat\xi_2 & = - \sigma_3 \otimes \sigma_3 \otimes \sigma_2 -i \sigma_3 \otimes \sigma_2 \otimes \sigma_2 \, ,\nonumber\\
\sqrt2\, \hat\xi_3 & = \sigma_3 \otimes \mathds{1} \otimes \sigma_3 + i \sigma_2 \otimes \sigma_1 \otimes \sigma_3\, , \nonumber
\end{align}
and the canonical momenta are given by
\begin{align}
2\sqrt2\, \hat\pi_1 & = i \mathds{1}\otimes \mathds{1} \otimes \sigma_1 - \mathds{1} \otimes \sigma_1\otimes \sigma_2\, , \nonumber\\
2\sqrt2\, \hat\pi_2 & = -i \sigma_3 \otimes \sigma_3 \otimes \sigma_2 - \sigma_3 \otimes \sigma_2 \otimes \sigma_2\, , \nonumber\\
2\sqrt2\, \hat\pi_3 & = i \sigma_3 \otimes \mathds{1} \otimes \sigma_3 +  \sigma_2 \otimes \sigma_1 \otimes \sigma_3\, . \nonumber
\end{align}
The Hanson-Regge-Teitelboim primed operators $\hat\xi^\prime_i$ are given by 
\begin{align}
\sqrt2\, \hat\xi^\prime_1 & = \mathds{1} \otimes \mathds{1} \otimes \sigma_1\, , \nonumber\\
\sqrt2\, \hat\xi^\prime_2 & = - \sigma_3 \otimes \sigma_3 \otimes \sigma_2\, , \nonumber\\
\sqrt2\,\hat\xi^\prime_3 & = \sigma_3 \otimes \mathds{1} \otimes \sigma_3\, , \nonumber
\end{align}
and the constraint operators $\hat\varphi_i$ are given by
\begin{align}
\sqrt2\, \hat\varphi_1 & = -\, \mathds{1} \otimes \sigma_1 \otimes \sigma_2\, , \nonumber\\
\sqrt2\, \hat\varphi_2 & = -\, \sigma_3 \otimes \sigma_2 \otimes \sigma_2\, , \nonumber\\
\sqrt2\, \hat\varphi_3 & =  \sigma_2 \otimes \sigma_1 \otimes \sigma_3\, . \nonumber
\end{align}
The canonical angular momentum components are
\begin{align}
\hat{S}_{23} & = \frac{1}{2}\left(-\mathds{1} \otimes \sigma_3 \otimes \sigma_1 + \sigma_1 \otimes \sigma_3 \otimes \sigma_1\right)\, ,\nonumber\\
\hat{S}_{31} & = \frac{1}{2}\left(\sigma_3 \otimes \mathds{1} \otimes \sigma_2 + \sigma_2 \otimes \mathds{1} \otimes \sigma_1\right)\, ,\nonumber\\
\hat{S}_{12} & = \frac{1}{2}\left(\sigma_3 \otimes \sigma_3 \otimes \mathds{1} - \sigma_3 \otimes \sigma_3 \otimes \sigma_3\right)\, ,\nonumber
\end{align}
and the spin generators are
\begin{align}
\hat\Sigma_{23} & =  -\, \frac{1}{2}\, \mathds{1} \otimes \sigma_3 \otimes \sigma_1\, , \nonumber\\
\hat\Sigma_{31} & =  \frac{1}{2}\, \sigma_3 \otimes \mathds{1} \otimes \sigma_2\, , \nonumber\\
\hat\Sigma_{12} & =   -\, \frac{1}{2}\, \sigma_3 \otimes \sigma_3 \otimes \sigma_3\, . \nonumber
\end{align}
The bilinear constraint operators $-\frac{i}{2} \hat\varphi_{[i} \hat\varphi_{j]}$ are
\begin{align}
-i\hat\varphi_{2}\hat\varphi_{3} & =  \frac{1}{2}\,\sigma_{1} \otimes \sigma_3 \otimes \sigma_1\, , \nonumber\\
-i\hat\varphi_{3}\hat\varphi_{1} & =  \frac{1}{2}\,\sigma_{2} \otimes \mathds{1} \otimes \sigma_1\, , \nonumber\\
-i\hat\varphi_{1}\hat\varphi_{2} & =  \frac{1}{2}\,\sigma_{3} \otimes \sigma_3 \otimes \mathds{1}\, . \nonumber
\end{align}

\section{Orthonormal basis of $\mathfrak{so}(n)$ irreps}

Here we exhibit an orthonormal basis for the physical states and ghost
states as antisymmetric irreducible representations of $\mathfrak{so}(n)$
built from the set of $2^n$ possible components of a function on the {\it
  n}-dimensional configuration space. To simplify the expressions, we use
the scaled variables, $\zeta^\mu = \xi^\mu/\sqrt2$, as suggested by the
form of the physical states\cite{Allen:2015lma} as given, for example, by
Eq.~(\ref{eq:threexiphys}) for $n=3$.

A basis of the full physical plus ghost state space is the set of all
unique monomials
\begin{equation}\label{eq:allmonomials}
\zeta^{\mu_1}\zeta^{\mu_2}\cdots\zeta^{\mu_k},  
\end{equation}
where $k$ runs from $0$ to $n$. When $k=0$ the monomial is defined to be
$1$. This basis is neither orthogonal nor normalized, nor are the states
purely physical or ghost, though they are either purely real or purely
imaginary.

Monomials with more than $\lfloor \frac{n}{2} \rfloor$ indices may be
labeled with $\lfloor\frac{n}{2}\rfloor$ indices or fewer by using the
$\mathfrak{so}(n)$ invariant tensor $\epsilon^{\mu_1 \mu_2 \cdots \mu_n}$
($\epsilon^{12\cdots n}=\epsilon_{12\cdots n}=1$) to define
\begin{equation}\label{eq:moreindices}
\textasteriskcentered(\zeta^{\mu_1} \cdots \zeta^{\mu_k}) = \frac{1}{(n-k)!}\epsilon_{\mu_1\cdots\mu_k\rho_1\cdots\rho_{n-k}} \zeta^{\rho_1}\zeta^{\rho_2}\cdots\zeta^{\rho_{n-k}}\, . 
\end{equation}

Orthonormal bases for the physical and ghost state spaces are the set of
unique linear combinations of monomials in Eqs.~(\ref{eq:allmonomials}) and
(\ref{eq:moreindices}) with at most $\lfloor\frac{n}{2} \rfloor$ indices
given by
\begin{equation}\label{eq:sonbasis}
|\mu_1,\mu_2,\cdots,\mu_k,\,\pm \rangle = \strut\alpha_k\left[\zeta^{\mu_1}\cdots\zeta^{\mu_k} \pm \beta_k\, \textasteriskcentered(\zeta^{\mu_1} \cdots \zeta^{\mu_k})\right]\, ,
\end{equation}
where 
\begin{equation}\label{eq:betak}
\beta_k = {i^{\lfloor \frac{n}{2}\rfloor}} (-1)^{\lfloor \frac{k}{2}\rfloor}\, ,
\end{equation}
and 
\begin{equation}
\alpha_k = 2^{\frac{n-2}{4}}\, .
\end{equation}
With the positive sign, the set of unique states of form
(\ref{eq:sonbasis}) is an orthonormal basis for the physical states. With
the negative sign, that set is an (anti-)orthonormal basis of the ghost
states. For $n=2$, these states are given in Eqs.~(\ref{eq:fullphysbasis})
and (\ref{eq:fullghostbasis}) respectively, while for $n=1$, these two
states are
\begin{equation}
|\pm \rangle = \frac{1}{\sqrt[4]{2}} \left(1 \pm \zeta \right) .
\end{equation}

For even $n=2m$, the physical (or ghost) states with $m$ indicies satisfy
duality relations
\begin{equation}
| \mu_1,\cdots,\mu_m,\, \pm\rangle = \pm \frac{\beta_m}{m!} \epsilon^{\mu_1\cdots\mu_m\rho_1\cdots\rho_m}\, | \rho_1,\cdots,\rho_m,\,\pm \rangle \, .
\end{equation}

The set of unique states (\ref{eq:sonbasis}), those with $\mu_1 < \mu_2 <
\ldots < \mu_k$, for a given $k$ forms an irreducible representation of
$\mathfrak{so}(n)$, which can be seen by acting the Noether angular
momentum generators (\ref{eq:NoetherSpin}) on these states, which either
produce zero or another state of the same form.

It is immediate that the unique states of form (\ref{eq:sonbasis}) with
different $k$ are orthogonal under the inner product (\ref{eq:SchrodNorm})
and that the set of all unique states of form (\ref{eq:sonbasis}) is
linearly independent.

The number unique states of form (\ref{eq:sonbasis}) is $2^n$, half of them
physical and half of them ghost.  Thus we obtain the result that they must
form a basis for both the physical and ghost states.

A useful identity for verifying Eq.~(\ref{eq:betak}) is
\begin{equation}
{\lfloor \frac{n}{2} \rfloor + \lfloor \frac{k}{2} \rfloor +\lfloor \frac{n-k}{2} \rfloor + (n-k)k} \equiv 1 \quad (\textrm{mod}~2),
\end{equation}
for integers $0\leq k \leq n$.


\end{document}